\documentclass[conf]{new-aiaa}
\usepackage[utf8]{inputenc}

\usepackage{graphicx}
\usepackage{amsmath}
\usepackage[version=4]{mhchem}
\usepackage{siunitx}
\usepackage{wasysym}
\usepackage{longtable,tabularx}
\usepackage{tikz}
\usepackage{multirow}
\DeclareMathOperator*{\argmin}{arg\,min} % thin space, limits underneath in displays
\graphicspath{{Figures/}} %Setting the graphicspath

\title{Neural representation of a time optimal, constant acceleration rendezvous}

\author{Dario Izzo\footnote{Scientific coordinator, Advanced Concepts Team, ESA-ESTEC} and Sebastien Origer\footnote{Scientist, Advanced Concepts Team, ESA-ESTEC} }
\affil{European Space Technology and Research Centre, 2201 AZ Noordwijk, The Netherlands}

\begin{document}

\maketitle

\begin{abstract}
    We train neural models to represent both the optimal policy (i.e. the optimal thrust direction) and the value function (i.e. the time of flight) for a time optimal, constant acceleration low-thrust rendezvous. In both cases we develop and make use of the data augmentation technique we call backward generation of optimal examples. We are thus able to produce and work with large dataset and to fully exploit the benefit of employing a deep learning framework. We achieve, in all cases, accuracies resulting in successful rendezvous (simulated following the learned policy) and time of flight predictions (using the learned value function). We find that residuals as small as a few m/s, thus well within the possibility of a spacecraft navigation $\Delta V$ budget, are achievable for the velocity at rendezvous. We also find that, on average, the absolute error to predict the optimal time of flight to rendezvous from any orbit in the asteroid belt to an Earth-like orbit is small (less than 4\%) and thus also of interest for practical uses, for example, during preliminary mission design phases.
\end{abstract}

\section*{Introduction}
\lettrine{D}{eep} neural models for optimal spacecraft guidance and control have been recently proposed \cite{sanchez2018real, dario_ekin_earth_venus, li2019neural, federici2021deep, hovell2020deep, cheng2018real} and studied as a possible alternative to more classical architectures.
Previously, researchers had proposed the use of \lq\lq neurocontrollers\rq\rq\ \cite{mohammadzaheri2012critical} as part of the guidance and control system, but mostly using a neural model to track some precomputed reference guidance signal. More recent works succeeded in learning directly the solution to Bellman's equations \cite{bardi1997optimal} hence synthesising an optimal state feedback for the underlying nonlinear equations of motion and thus allowing the use of networks for guidance and control simultaneously.
These new types of networks, which we call Guidance and Control Networks (G\&CNETs), have a modest computational cost during inference, with clear applications for on-board, autonomous guidance and control \cite{dario_ekin_earth_venus, sanchez2018real, tailor2019learning}. Training them, on the other hand, is mostly done via supervised learning and requires the generation and use of a great number of optimal trajectories to be used as labels to learn from. The generation of a suitable dataset is key to the success in training any neural model to achieve the precision requested by real mission design constraints. To this end, an indirect method based on Pontryagin's approach \cite{pontryagin, jiang2012practical} can be used to generate (and later learn from) and label optimal decisions. As a consequence, the solution to many Two Points Boundary Value Problems (TPBVPs) needs to be computed. Even in the optimistic case of good co-states guesses being available \cite{li2019autonomous, li2019neural} and homotopy methods employed, these numerical procedures are time consuming and limit the amount of reference trajectories one can learn from \cite{cheng2018real, li2019neural}. A similar conclusion can be made also when direct methods are used to generate optimal examples.
The development of neural models representing the optimal policy of some interplanetary transfer, such as G\&CNETs, is not the only case where the availability of a large amount of optimal solutions is needed.
Attempts to learn directly machine learning models (mostly  deep neural models) for the propellant consumption, the $\Delta V$, the arrival mass or the time of flight have also been performed. These can all be seen, from an optimal control point of view, as attempts to learn with different degrees of approximation the value function of some underlying fundamental optimal control problem.
Complex low-thrust interplanetary trajectories between objects of the main belt \cite{hennes2016fast, li2020deep}, transfers to Earth co-orbiting asteroids \cite{mereta2017machine} as well as orbiting debris scenarios\cite{shen2021simple} have been considered as applications in the past.
A critic to these approaches for value function learning has been recently worded as: \lq\lq ... one has to balance the advantage of computation speed against a remarkable time for data sets generating and network training.\rq\rq \cite{shen2021simple}.

In this work we utilise a constant acceleration interplanetary rendezvous problem, to show the use of the data augmentation technique called Backward Generation of Optimal Examples \cite{dario_ekin_earth_venus} as a methodology to reduce the time required to generate datasets by orders of magnitude.
The method allows to generate optimal samples avoiding entirely the solution of TPBVPs by simply integrating backward in time the augmented system of differential equations derived from Pontryagin's theory. 
We then also show how, contrary to a common sentiment, accuracies compatible with real world requirements can be reached both when learning the optimal policy and the value function.

The paper is structured as follows. 
We begin giving the necessary details and formalism on the optimal control problem considered throughout the paper.
In doing so we also follow recent practical guidelines \cite{jiang2012practical} for solving low-thrust problems via the maximum principle, which reveal to have added specific advantages for the augmentation technique.
We then show how to apply the backward generation of optimal examples to augment a single nominal optimal trajectory found and discuss the results obtained training several G\&CNETs for different initial orbits.
In the following section we move on to consider the value function problem and discuss the results one can obtain when augmenting each trajectory in a dataset aimed at representing the time of transfer from orbits in the asteroid belt to a target rendezvous.

\section*{Methods}
\subsection*{The optimal control problem}
\label{sec:ocp}
Consider a time optimal, constant acceleration rendezvous with a body in a perfectly circular orbit. This problem was recently brought to the attention of a wide international community during the 11th edition of the Global Trajectory Optimisation Competition (GTOC11) where it represented low-thrust transfers of objects from the belt. Let us introduce a rotating frame $\mathcal F = \left[\hat{\mathbf i}, \hat{\mathbf j}, \hat{\mathbf k}\right]$ having angular velocity $\boldsymbol \omega= \sqrt{\frac{\mu}{R^3}} \hat {\mathbf k}$ so that the target body position is, in $\mathcal F$,  stationary and indicated with $R\hat{\mathbf i}$. 
The dynamics is thus described by the following simple set of ordinary differential equations:
\begin{equation}
\label{eq:dyn}
\left\{ 
\begin{array}{l}
    \dot{\mathbf{r}} = \mathbf{v} \\
    \dot{\mathbf{v}} = -\frac{\mu}{r^3}\mathbf{r}-2\boldsymbol \omega\times\mathbf v-\boldsymbol \omega\times\boldsymbol \omega\times\mathbf r +\Gamma\mathbf{\hat{t}}(t)
\end{array}
\right.
\end{equation}
where $\mu$ is the gravitational constant of the Sun and  $\Gamma$ a constant acceleration assumed to be acting along some direction (to be controlled) described by the unit vector $\mathbf{\hat{t}}$. We will consider, for the rest of this work, $\Gamma = 0.1$ mm/s$^2$ which is a value compatible with successful interplanetary low-thrust missions such as Dawn \cite{russell2011dawn}, Bepi-Colombo \cite{benkhoff2010bepicolombo} or Smart-1 \cite{foing2006smart}. We consider the problem of finding $t_f$ and a (piece-wise continuous) function $\mathbf{\hat{t}}(t)$ with $t \in [t_0, t_f]$ so that, under the dynamics defined in Eq.(\ref{eq:dyn}), the state is steered from any initial state $\mathbf r_0, \mathbf v_0$ to the target state $\mathbf r_t = R\hat{\mathbf i}, \mathbf v_t = \mathbf 0$ in a minimal amount of time, thus minimizing the following cost function:
$J = t_f-t_0 = \int^{t_f}_{t_0}  dt$. Following Pontryagin's Minimum Principle \cite{pontryagin} and some of the practical guidelines suggested by Jiang and Baoyin \cite{jiang2012practical} we introduce the following Hamiltonian: 
\begin{equation}
\label{eq:hamiltonian}
    \mathcal H(\mathbf{r},\mathbf{v},\boldsymbol{\lambda}_r,\boldsymbol{\lambda}_v,\mathbf{\hat{t}})= \boldsymbol{\lambda}_r\cdot \mathbf{v}+\boldsymbol{\lambda}_v\cdot\bigg(-\frac{\mu}{r^3}\mathbf{r}-2\boldsymbol \omega\times\mathbf v-\boldsymbol \omega\times\boldsymbol \omega\times\mathbf r+\Gamma\mathbf{\hat{t}}\bigg)+\lambda_J
\end{equation}
where we have introduced the co-states functions $\mathbf{\lambda_r}$ and $\mathbf{\lambda_v}$ and an additional $\lambda_J$ constant coefficient that multiplies the actual cost function considered $J=\lambda_J (t_f-t_0)$. The additional coefficient $\lambda_J$ results in increased numerical stability and, as we shall see, offers a useful extra degree of freedom for data augmentation.

Since the Hamiltonian must be minimized by the optimal control (i.e. $\mathbf{\hat{t}}^*$) during an optimal trajectory, it is straight forward to derive the classical necessary condition:
$$
\mathbf{\hat{t}}^*=-\frac{ \boldsymbol{\lambda}_{\mathbf v}}{\lambda_v}
$$
The augmented dynamics is then derived from the Hamiltonian performing the derivatives $\dot {\mathbf x} = \frac{\partial \mathcal H}{\partial \mathbf x}$, $\dot {\boldsymbol \lambda} = - \frac{\partial \mathcal H}{\partial \boldsymbol \lambda}$:
\begin{equation}
\label{eq:augdyn}
\left\{ 
\begin{array}{l}
    \dot{\mathbf{r}} = \mathbf{v} \\
    \dot{\mathbf{v}} =  -\frac{\mu}{r^3}\mathbf{r}-2\boldsymbol \omega\times\mathbf v-\boldsymbol \omega\times\boldsymbol \omega\times\mathbf r-\Gamma\frac{ \boldsymbol{\lambda}_{\mathbf v}}{\lambda_v}\\
    \dot{\boldsymbol\lambda}_{\mathbf{r}} = \mu \left(\frac{\boldsymbol{\lambda}_{\boldsymbol v}}{r^3} - 3(\boldsymbol \lambda_{\boldsymbol v}\cdot\mathbf r)\frac{\mathbf r}{r^5} \right) - \boldsymbol\omega\times\boldsymbol\omega\times\boldsymbol \lambda_v\ \\
    \dot{\boldsymbol\lambda}_{\mathbf{v}} = - \boldsymbol{\lambda}_{\mathbf r} +2\boldsymbol\omega\times\boldsymbol\lambda_v
\end{array}
\right.
\end{equation}
According to Pontryagin's theory, the optimal transfer will necessarily be a solution to the above differential equations with the added condition $\mathcal H=0$ as we consider a free time problem. The inverse is also true: any solution to the above system of differential equations, respecting the added condition on the Hamiltonian, is an optimal trajectory and can thus be used to learn from, regardless of where it actually starts from. A point of view that, as we shall see, is at the basis of the backward generation of optimal examples here used to form the database of state action pairs.

\begin{figure}[t]
\centering
\includegraphics[width=0.8\textwidth]{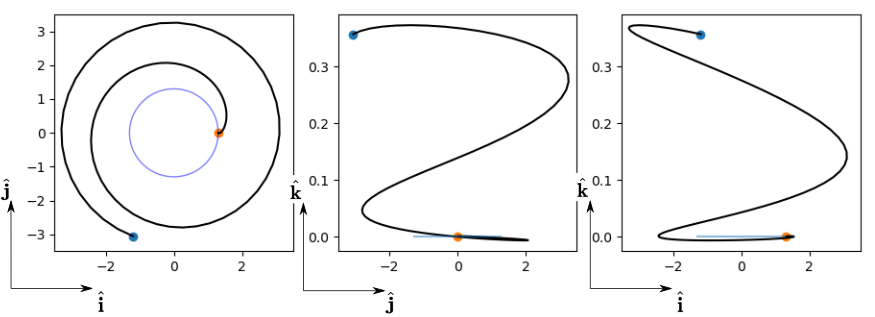}
\caption{Nominal trajectory used for policy learning}
\label{fig:nominal}
\end{figure}

\subsection*{Finding one optimal trajectory}
In order to find one solution to the optimal control problem outlined in the previous section, we introduce the following shooting function:
\begin{equation}
\label{eq:shooting}
\phi(\boldsymbol \lambda_{r_0}, \boldsymbol \lambda_{v_0}, \lambda_J, t_f) = \left\{\mathbf r(t_f) - \mathbf r_t,  \mathbf v(t_f) - \mathbf v_t, \mathcal H,||\boldsymbol \lambda || -1 \right\}
\end{equation}
where $\mathbf r(t_f)$ and $\mathbf v(t_f)$ (and $\mathcal H$) are obtained by numerically propagating Eq.(\ref{eq:augdyn}) from the assigned initial conditions $\mathbf r_0, \mathbf v_0, \boldsymbol \lambda_{r_0}, \boldsymbol \lambda_{v_0}$ for a time $t_f$. The shooting function is a system of eight non-linear relations between the variables $\boldsymbol \lambda_{r_0}, \boldsymbol \lambda_{v_0}, \lambda_J, t_f$: its root is here found numerically applying the sequential quadratic programming solver SNOPT \cite{gill2005snopt} and defines entirely an optimal trajectory. Note that the condition on the initial co-state norm is not strictly necessary in this case, but it helps with the numerical stability of the root finder.

As an example, let us here consider an orbit having, at epoch $t_0$, the following osculating orbital Keplerian elements in the inertial frame: $a = 2.687$ AU, $e = 0.23$, $i = 0.116$ rad., $\Omega = 3.137$ rad., $\omega = 4.453$ rad. and eccentric anomaly $E = 3.01$ rad. 
Without loss of generality we assume that, at $t_0$, the rotating frame is coincident to the inertial frame, so that a rotation matrix between the two frames will take the simple form:
$$
\mathbf R(t)=
\left[
\begin{array}{ccc}
\cos{\omega (t-t_0)} & \sin {\omega (t-t_0)} & 0 \\
-\sin{\omega (t-t_0)} & \cos {\omega (t-t_0)} & 0 \\
0 & 0& 1 
\end{array}
\right]
$$
Given the Keplerian osculating elements at $t$ in the inertial frame we thus 
find the initial conditions $\mathbf r_0, \mathbf v_0$ in the rotating frame $\mathcal F$ and find the root of Eq.(\ref{eq:shooting}) when the target orbit is at $R=1.3$ AU. 
The resulting trajectory is shown in Figure \ref{fig:nominal} and has a total time of flight of $t_f^*=4.62$ years. Being a solution to the shooting equation is a necessary condition for optimality but not a sufficient one. As a consequence, there may be multiple roots for the shooting equation corresponding to different transfer strategies. This corresponds to the existence of multiple local minima and is a problem hard to tackle rigorously within Pontryagin's theory. Typically these multiple roots can be found restarting the root finder from different initial guesses for the co-states, an approach that, although not rigorous, is often taken and deemed as unavoidable. For this particular nominal trajectory we verified experimentally that no other root exists via multiple restarts, but as we shall see later when assembling a large dataset of nominal trajectories, it is challenging to guarantee mathematically that this is the case.

\subsection*{Augmenting one optimal trajectory}
\label{sec:augmenting}
To create a dataset of optimal trajectories one typically would solve multiple times the shooting function in Eq.(\ref{eq:shooting}) using different values of $\mathbf r_0$ and $\mathbf v_0$. In this way, a significant amount of computational resources are needed. In more difficult cases, where a homotopy technique is necessary (see \cite{bertrand2002new} for an example) and local minima and convergence are issues, it may just be impossible to compute the number of optimal trajectories required to train an accurate model. A far more efficient method is to use the Backward Generation of Optimal Examples \cite{dario_ekin_earth_venus}. This technique can be seen as a data augmentation technique where each sample in a dataset of optimal trajectories can be used, at low computational cost, to generate many more different samples. In order to apply this technique to the particular case at hand, we indicate the final values of the co-states  $\boldsymbol \lambda_r$, $\boldsymbol \lambda_v$ relative to one optimal trajectory with $\boldsymbol \lambda_f$. We thus introduce the update rule:

\begin{equation}\label{eq:pert}
   \boldsymbol \lambda_{f}^{+} = \boldsymbol \lambda_{f} + \boldsymbol \lambda_{f}\cdot\boldsymbol\Delta
\end{equation}
Where $\boldsymbol\Delta$ is a perturbation vector chosen as to satisfy all of the conditions dictated by Pontryagin's principle, in our case only $\mathcal H_f = 0$. In this work we will often use for each element of the perturbation vector a uniformly distributed number $\mathcal U(-\delta, \delta)$.
The benefit of having introduced the normalisation factor $\lambda_J$ appears now clearly.
In fact, setting Eq.(\ref{eq:hamiltonian}) to zero, we see how we are allowed to choose all of the perturbations on the co-states freely since we can always set $\lambda_J^{new}$ as follow:
$$
\lambda_J^{+} = -\boldsymbol{\lambda}_{\mathbf r_f}^{+}\cdot \mathbf{v}_f-\boldsymbol{\lambda}_{\mathbf v_f}^{+}\cdot\bigg(-\frac{\mu}{r_f^3}\mathbf{r}_f-2\boldsymbol \omega\times\mathbf v_f-\boldsymbol \omega\times\boldsymbol \omega\times\mathbf r_f-\Gamma\frac{
\boldsymbol{\lambda}_{\mathbf v_f}^{+}}{{\lambda}_{v_f}^{+}}\bigg)
$$
and thus guarantee, for any arbitrary perturbation vector, that $\mathcal H_f = 0$. 
Note that with respect to the case treated in \cite{dario_ekin_earth_venus} we have here simplified the overall complexity of the dataset generation further by avoiding the solution of a nonlinear equation representing the transversality condition on the Hamiltonian. 
Once a new valid value $\boldsymbol \lambda_{f}^{+}$ for the final co-states is available we propagate backward Eq.(\ref{eq:augdyn}) and thus obtain an optimal trajectory to be added to the dataset. 
In this work we use a Taylor technique \cite{biscani2021revisiting} to perform a highly accurate numerical propagation and we are able to generate 10,000 optimal trajectories in the order of seconds: orders of magnitude less than directly solving the corresponding shooting function. Of course, it must be noted, the exact initial conditions of all these optimal trajectories are not chosen, but a consequence of the new final co-states values. 

\section*{Neural representation of the optimal policy}
In this section we train a neural model $\mathcal N$ to represent the optimal policy $\mathbf{\hat{t}}^*$ as a function of the spacecraft state $\mathbf r, \mathbf v$. The resulting neural state feedback $\mathbf{\hat{t}}^* = \mathcal N(\mathbf r, \mathbf v)$, called a G\&CNET, is then used in Eq.(\ref{eq:dyn}) to simulate the spacecraft dynamics.

\subsection*{The dataset}
\begin{figure}[tb]
\centering
\includegraphics[width=0.9\textwidth]{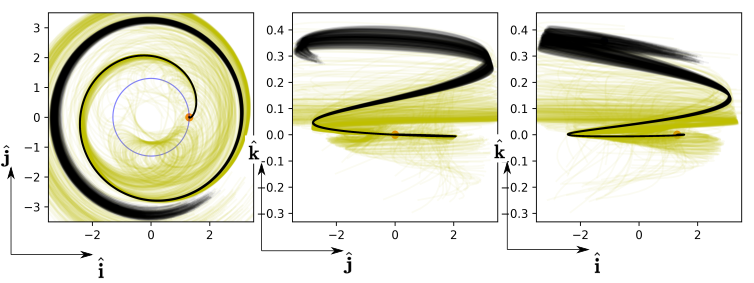}
\caption{Bundle of ~200 optimal trajectories obtained applying the backward generation of optimal examples. Co-state perturbation is 1 \permil (black), 8\% (yellow).}
\label{fig:bundle}
\end{figure}
We generate a dataset of optimal state-action pairs augmenting one single nominal trajectory via the Backward Generation of Optimal Examples. The generic single entry in our dataset thus contains the values $[X_i,Y_i] = (x_i,y_i,z_i,{v_x}_i,{v_y}_i,{v_z}_i),({t_x}_i^*,{t_y}_i^*,{t_z}_i^*)$. These are computed generating 200,000 optimal trajectories sampled in 100 points equally distanced in time. The resulting dataset thus contains 20,000,000 optimal state-action pairs.
Half of the dataset is generated perturbing the final co-states $\lambda_f^{+}$ using $\delta = 0.001$ and back-propagating the augmented equations of motion for a time $(1+c)t^*_f$ with the constant $c$ uniformly distributed  $c\in[0,1]$. 
The resulting bundle, shown in Figure \ref{fig:bundle}, grows very thin as the spacecraft approaches the final rendezvous. As a result, small deviations from the nominal trajectory may lead, during a real transfer, the spacecraft state to be outside of the training data and the resulting error to grow to unacceptable levels. 
To overcome this issue, the remaining half of the dataset is generated using a larger perturbation $\delta = 0.08$.
The resulting dataset is visualised in Figure \ref{fig:bundle}, where the different colours correspond to the two halves of the dataset. We note how the larger perturbation leads to trajectories that may also depart significantly from the nominal trajectory. Interestingly, during our experiments, we found these trajectories important to a successful training procedure, likely as they thicken the bundle during the second half of the interplanetary transfer. 
The CPU time required to assemble the entire dataset, on a standard modern CPU (Intel\textregistered\ Core\texttrademark\ i5-6600K CPU, 3.50 GHz), is less than one minute. 
We also generate a completely independent test set containing 4,000 whole trajectories computed using $\delta = 0.0005$ and back-propagating the augmented equations of motion for a time $(1+c)t^*_f$. The smaller perturbation vector used to generate the test set is aimed at making sure to only probe the model within the bounds of the training data. It is here worth mentioning, though, that we also used and generated a test set with the same perturbation vector as that used to generate the training set and obtained, essentially, similar results.
All numerical integrations are performed using Heyoka \cite{biscani2021revisiting}, a Taylor based method, and setting the numerical tolerance to machine level, i.e. $\epsilon = 1\cdot 10^{-16}$. 

\begin{figure}[t]
\centering
\tikzset{%
  every neuron/.style={
    circle,
    draw,
    minimum size=0.6cm
  },
  neuron missing/.style={
    draw=none, 
    scale=4,
    text height=0.333cm,
    execute at begin node=\color{black}$\vdots$
  },
}

\begin{tikzpicture}[x=1.5cm, y=1cm, >=stealth]

% Nodes
\foreach \m/\l [count=\y] in {1,2,3,4,5,6}
  \node [every neuron/.try, neuron \m/.try] (input-\m) at (0,2.5-\y) {};

\foreach \m [count=\y] in {1,missing,2}
  \node [every neuron/.try, neuron \m/.try ] (hidden-\m) at (1.3,1.5-\y*1.25) {};
  
\foreach \m [count=\y] in {1,missing,2}
  \node [every neuron/.try, neuron \m/.try ] (hidden2-\m) at (2.6,1.5-\y*1.25) {};
  
 \foreach \m [count=\y] in {1,missing,2}
  \node [every neuron/.try, neuron \m/.try ] (hidden3-\m) at (3.9,1.5-\y*1.25) {};
  
  \foreach \m [count=\y] in {1,missing,2}
  \node [every neuron/.try, neuron \m/.try ] (hidden4-\m) at (5.2,1.5-\y*1.25) {};

\foreach \m [count=\y] in {1,2,3}
  \node [every neuron/.try, neuron \m/.try ] (output-\m) at (6.5,1-\y) {};

% Annotations
\foreach \l [count=\i] in {x,y,z,v_x,v_y,v_z}
  \draw [<-] (input-\i) -- ++(-1,0)
    node [above, midway] {$\l$};

% \foreach \l [count=\i] in {1,700}
%   \node [above] at (hidden-\i.center) {${\l}$};
  
%  \foreach \l [count=\i] in {1,700}
%   \node [above] at (hidden2-\i.south) {$H_{\l}$};

\foreach \l [count=\i] in {x,y,z}
  \draw [->] (output-\i) -- ++(1,0)
    node [above, midway] {$t_\l$};

% Connections
\foreach \i in {1,2,3,4,5,6}
  \foreach \j in {1,...,2}
    \draw [->] (input-\i) -- (hidden-\j);

\foreach \i in {1,...,2}
  \foreach \j in {1,...,2}
    \draw [->] (hidden-\i) -- (hidden2-\j);

\foreach \i in {1,...,2}
  \foreach \j in {1,...,2}
    \draw [->] (hidden2-\i) -- (hidden3-\j);
    
\foreach \i in {1,...,2}
  \foreach \j in {1,...,2}
    \draw [->] (hidden3-\i) -- (hidden4-\j);

\foreach \i in {1,...,2}
  \foreach \j in {1,2,3}
    \draw [->] (hidden4-\i) -- (output-\j);

% Titles
\foreach \l [count=\x from 1] in {(Softplus) \\ Hidden, (Softplus) \\ Hidden,(Softplus) \\ Hidden,(Softplus) \\ Hidden, (Linear) \\ Output}
  \node [align=center, above] at (\x*1.3,2) {\l \\ layer};

\end{tikzpicture}
\caption{Feed forward neural network architecture}

\label{fig:ffnn}
\end{figure}
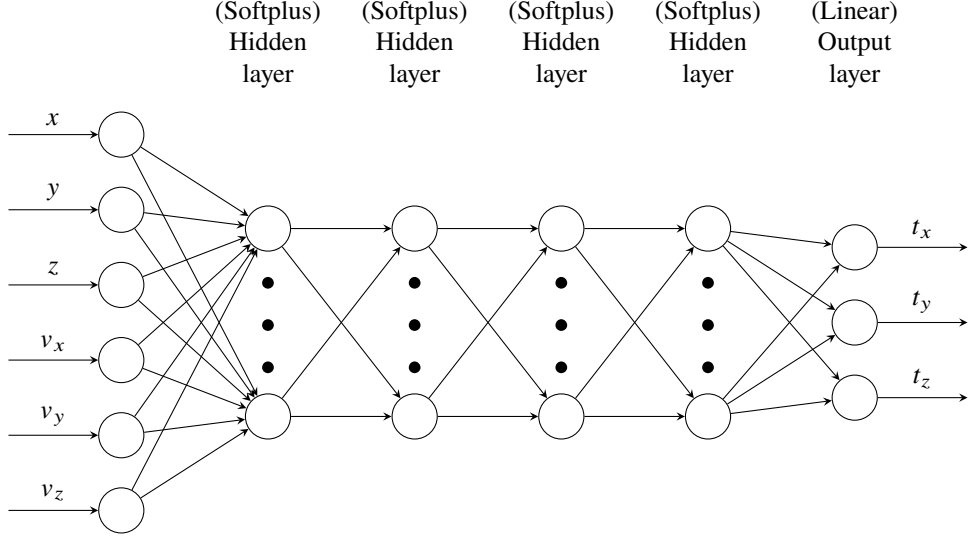

\subsection*{The neural model}
We consider learning the optimal thrust from the spacecraft states as a supervised learning task and we make use of a simple feed forward neural network, as introduced in \cite{sanchez2018real, dario_ekin_earth_venus} for some related tasks. 
A softplus activation function is used for the input and the hidden layers, allowing to obtain a continuous representation of the optimal controls differentiable to high order \cite{tailor2019learning}. 
We use four hidden layers of 700 neurons each. For the output layer we use a linear activation function, avoiding possible saturation issues. 
The network inputs are the states ($x$, $y$, $z$, $v_x$, $v_y$, $v_z$) and the outputs are the thrust direction ($t_x$, $t_y$, $t_z$), see Figure~\ref{fig:ffnn}. We will also use the notation $\mathcal N(\mathbf r, \mathbf v) = \hat{\mathbf t}_{nn}$ to indicate the network predictions.

As a loss function we make use of the cosine similarity between the estimated thrust direction vector $\mathbf{\hat{t}}_{nn}$ and the ground truth $\mathbf{\hat{t}}^*$. 
In particular, the cosine similarity is defined as:
$$
\text{cosine\_similarity}\left(\mathbf{\hat{t}}^*, \mathbf{\hat{t}}_{nn}\right) = \frac{\mathbf{\hat{t}}^*\cdot \mathbf{\hat{t}}_{nn}}{\mathbf{\vert\hat{t}}^*\vert\vert \mathbf{\hat{t}}_{nn}\vert}
$$
and the loss used:
$$
\mathcal L\left(\mathbf{\hat{t}}_{nn},\mathbf{\hat{t}}^*\right) = 1-\text{cosine\_similarity}\left(\mathbf{\hat{t}}_{nn}, \mathbf{\hat{t}}^*\right)
$$
This formulation of the loss function allows the network to focus on predicting the thrust direction only, ignoring any effect of the predicted norm. 
Together with the last linear layer used, this allows for an efficient training. We use $80\%$ of the dataset for training, and the remaining $20\%$ for validation. 
The network weights are learned using the Adam optimizer \cite{kingma2014adam} with a starting learning rate $l=0.5\cdot 10^{-4}$ and no weight decay.
A learning rate schedule was imposed by reducing the rate by a factor $f=0.9$ after $p=10$ consecutive epochs without any improvement on the loss as computed on a random batch of the validation data. 
Numerical experiments were performed with other losses as well as alternative representations of the thrust direction, revealing, in our case, this setup to be the most accurate.

\subsection*{Results and discussion}

\begin{figure}[tb]
\centering
\includegraphics[width=\textwidth]{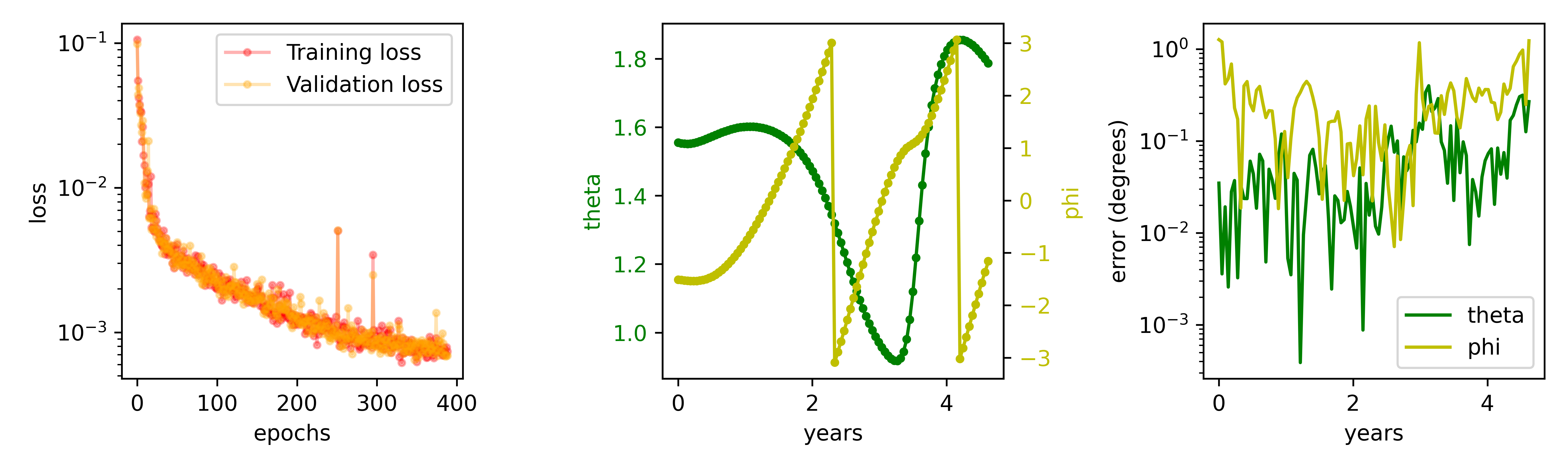}
\caption{Loss during a typical training (left). Networks predictions vs. ground truth during one optimal trajectory in the test set (centre and right).}
\label{fig:loss}
\end{figure}
\noindent
We train several neural models following the methodology outlined in the previous sections.
In Figure \ref{fig:loss} we report the loss decrease during a typical training, as well as an example of the predictions made by the neural network on a previously unseen optimal trajectory from the test set. 
In particular, the network prediction is transformed into spherical coordinates $\theta = \arccos{t_z}$ and $\varphi = \arctan{\frac{t_y}{t_x}}$ and compared to the optimal ground truth. 
The final loss value decreases below $10^{-3}$ and corresponds, as evaluated on the test set, to an average error of $0.59$ degrees in the prediction of the optimal direction. 
In order to assess if this error is small enough to allow the network to successfully steer the spacecraft optimally, we simulate the spacecraft dynamics in Eq.(\ref{eq:dyn}) when $\hat{\mathbf i}(t) = \mathcal N(\mathbf r(t), \mathbf v(t))$, that is when the thrust direction is instantaneously given by the network predictions.
We consider all initial conditions in the test set, numerically integrating the equations of motion from there.
We record the system state achieved at the time $t_f^*$ as well as at the time when the target semi-major axis $a_t$ is acquired exactly, at a time past $t_f^*$. 
The results, for the nominal trajectory used as test case in the previous sections, are given in Table \ref{tab:errors} (case N) and show how the spacecraft is able to rendezvous with the target body with good accuracy when the low-thrust profile follows the trained neural model. 
The network, acting as a state-feedback, is able to cancel out, rather than accumulate, the prediction errors possibly made during the transfer.
To confirm quantitatively the method accuracy and study its reliability, we repeat the very same experiment starting from 9 different nominal trajectories, identified with the capital letters A,B,..I, having all the same orbital elements, but a different initial eccentric anomaly and hence a difference phasing to the target. 
In all cases the trained G\&CNET is able to steer the spacecraft to the target rendezvous reliably and with similar accuracies. When integrating until $t_f^*$, the spacecraft is left within $~500,000$ Km and with a residual relative velocity of the order of $~100$ [m/s]. In the better cases (trajectories B, F) residuals of less than $~20$ m/s are reached on the final velocity and of the order of 100,000 km on the position. 
These residuals are compatible with an operational use of the network for automated on-board guidance and control. 
It has to be noted here that the numbers appearing in table \ref{tab:errors} have been assembled using the very same dataset generation and training parameters, e.g. $\delta$, $c$, $l$. 
Since the sensitivity of the  training accuracy to these parameters is rather significant, it is expected that during a real design phase one would be able to tune these parameters to the specific case (and dynamics) as to obtain a G\&CNET fulfilling operational constraints to the requested accuracies.

\begin{table}[tb]
    \centering
    \caption{Mean errors at target when propagating the neural dynamics. The numerical integration stops either at $t^*_f$ or when the spacecraft achieves the target orbital energy (i.e. semi major axis $a_t$).\label{tab:errors}}
    \begin{tabular}{c|c|c|cccccc}
Case &   $E_0$, rad & &$\vert\vert\boldsymbol \delta \mathbf r\vert\vert$, AU & $\vert\vert\boldsymbol \delta \mathbf v\vert\vert$, km/s & $\delta a$, AU & $\delta e$ & $\delta i$, deg&$\delta t$, years\\\hline \hline

\multirow{2}{*}{N} &  \multirow{2}{*}{$3.01 $}   & stop at $t_f^*$&
  $1.15 \cdot 10^{-3}$ &$7.65 \cdot 10^{-2}$&$7.60\cdot 10^{-4}$&$2.88 \cdot 10^{-3}$&$1.57 \cdot 10^{-2}$ & 0 \\ 
 &    & stop at $a_t$&
  $1.16 \cdot 10^{-3}$ &$7.46 \cdot 10^{-2}$&0&$2.77 \cdot 10^{-3}$&$1.60 \cdot 10^{-2}$ & $2.52 \cdot 10^{-3}$\\
 \hline\hline

\multirow{2}{*}{A} &  \multirow{2}{*}{$0.0 $}   & stop at $t_f^*$&
  $1.48 \cdot 10^{-3}$ &$4.65 \cdot 10^{-2}$&$3.16 \cdot 10^{-5}$&$  1.55\cdot 10^{-3}$&$ 1.13 \cdot 10^{-2}$ & 0 \\ 
 &    & stop at $a_t$&
  $1.49 \cdot 10^{-3}$ &$4.68 \cdot 10^{-2}$&0&$ 1.56\cdot 10^{-3}$&$ 1.14\cdot 10^{-2}$ & $ 4.87\cdot 10^{-5}$\\
 \hline

\multirow{2}{*}{B} & \multirow{2}{*}{$0.698 $}    & stop at $t_f^*$&
  $5.09 \cdot 10^{-4}$ &$1.68 \cdot 10^{-2}$&$1.39 \cdot 10^{-4}$&$5.63  \cdot 10^{-4}$&$1.79  \cdot 10^{-2}$ & 0 \\ 
 &    & stop at $a_t$&
  $4.94 \cdot 10^{-4}$ &$1.66 \cdot 10^{-2}$&0&$5.57 \cdot 10^{-4}$&$1.79 \cdot 10^{-2}$ & $4.98 \cdot 10^{-4}$\\
 \hline

 \multirow{2}{*}{C} & \multirow{2}{*}{$1.396 $ }   & stop at $t_f^*$&
  $2.73 \cdot 10^{-3}$ &$1.58 \cdot 10^{-1}$&$2.27 \cdot 10^{-3}$&$5.93  \cdot 10^{-3}$&$7.98  \cdot 10^{-2}$ & 0 \\ 
 &    & stop at $a_t$&
  $2.76 \cdot 10^{-3}$ &$1.48 \cdot 10^{-1}$&0&$5.43 \cdot 10^{-3}$&$8.23 \cdot 10^{-2}$ & $7.60 \cdot 10^{-3}$\\
 \hline

 \multirow{2}{*}{D} & \multirow{2}{*}{$2.094 $}    & stop at $t_f^*$&
  $3.50 \cdot 10^{-3}$ &$1.50 \cdot 10^{-1}$&$2.20 \cdot 10^{-3}$&$5.87  \cdot 10^{-3}$&$ 2.37 \cdot 10^{-2}$ & 0 \\ 
 &    & stop at $a_t$&
  $ 3.41\cdot 10^{-3}$ &$1.44 \cdot 10^{-1}$&0&$5.40 \cdot 10^{-3}$&$2.37\cdot 10^{-2}$ & $7.11 \cdot 10^{-3}$\\
 \hline

 \multirow{2}{*}{E} & \multirow{2}{*}{$2.793 $}    & stop at $t_f^*$&
  $1.81 \cdot 10^{-3}$ &$8.94 \cdot 10^{-2}$&$8.77 \cdot 10^{-4}$&$3.39  \cdot 10^{-3}$&$1.04  \cdot 10^{-2}$ & 0 \\ 
 &    & stop at $a_t$&
  $1.84 \cdot 10^{-3}$ &$8.79 \cdot 10^{-2}$&0&$3.30 \cdot 10^{-3}$&$1.08 \cdot 10^{-2}$ & $ 2.96 \cdot 10^{-3}$\\
 \hline

 \multirow{2}{*}{F} &  \multirow{2}{*}{$3.491$}   & stop at $t_f^*$&
  $1.21 \cdot 10^{-3}$ &$1.56 \cdot 10^{-2}$&$2.09 \cdot 10^{-4}$&$5.57  \cdot 10^{-4}$&$6.40  \cdot 10^{-3}$ & 0 \\ 
 &    & stop at $a_t$&
  $1.15 \cdot 10^{-3}$ &$1.49 \cdot 10^{-2}$&0&$5.0 \cdot 10^{-4}$&$6.12 \cdot 10^{-3}$ & $7.21  \cdot 10^{-4}$\\
 \hline

 \multirow{2}{*}{G} &  \multirow{2}{*}{$4.189$}  & stop at $t_f^*$&
  $6.46 \cdot 10^{-4}$ &$3.54 \cdot 10^{-2}$&$4.90 \cdot 10^{-4}$&$1.17  \cdot 10^{-3}$&$7.02  \cdot 10^{-3}$ & 0 \\ 
 &    & stop at $a_t$&
  $6.31 \cdot 10^{-4}$ &$3.18 \cdot 10^{-2}$&0&$9.69 \cdot 10^{-4}$&$5.14 \cdot 10^{-3}$ & $1.51  \cdot 10^{-3}$\\
 \hline

\multirow{2}{*}{H} & \multirow{2}{*}{$4.887$}   & stop at $t_f^*$&
  $1.28 \cdot 10^{-3}$ &$5.22 \cdot 10^{-2}$&$2.49\cdot 10^{-4}$&$2.0 \cdot 10^{-3}$&$8.48 \cdot 10^{-3}$ & 0 \\ 
 &    & stop at $a_t$&
  $1.28 \cdot 10^{-3}$ &$5.19 \cdot 10^{-2}$&0&$1.98 \cdot 10^{-3}$&$8.52 \cdot 10^{-3}$ & $7.99\cdot 10^{-4}$\\
 \hline

\multirow{2}{*}{I} &  \multirow{2}{*}{$5.585 $}   & stop at $t_f^*$&
  $3.18 \cdot 10^{-3}$ &$1.41 \cdot 10^{-1}$&$6.02 \cdot 10^{-4}$&$3.78  \cdot 10^{-3}$&$3.64  \cdot 10^{-2}$ & 0 \\ 
 &    & stop at $a_t$&
  $3.20 \cdot 10^{-3}$ &$1.42 \cdot 10^{-1}$&0&$3.81 \cdot 10^{-3}$&$3.77 \cdot 10^{-2}$ & $2.11  \cdot 10^{-3}$\\
 \hline

    \end{tabular}
\end{table}

\section*{Neural representation of the value function}
%\begin{figure}[tb]
%\centering
%\includegraphics[width=0.6\textwidth]{error_vs_ampl_fac.png}
%\caption{Mean error [days] versus amplification factor over 1000 test trajectories}
%\label{fig:err_vs_ampl_fac}
%\end{figure}
\begin{figure}[tb]
\centering
\includegraphics[width=0.9\textwidth]{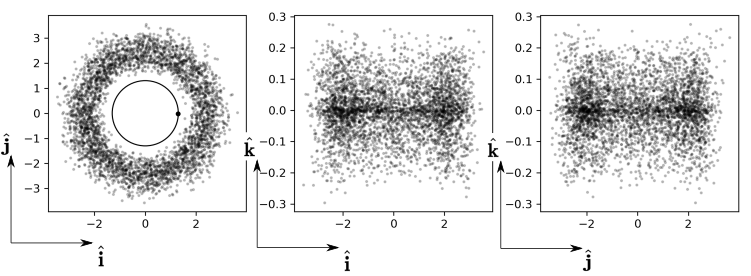}
\caption{Initial conditions in the value function learning dataset. Astronomical units used.}
\label{fig:belt}
\end{figure}
In this section we train a neural model $\mathcal N$ to represent the value function, or cost-to-go for our problem, that is $v(\mathbf r, \mathbf v) = J = t_f-t_0$: the optimal time of flight as a function of the state. 
Note that, in theory, if we were able to learn the value function exactly, the optimal policy could also be derived from Hamilton-Jacobi-Bellman (HJB) equations as:
$$
\mathbf t^* = \argmin_{|\mathbf t|=1}(\nabla_{\mathbf r} v \cdot \dot{\mathbf r} + \nabla_{\mathbf v} v \cdot \dot{\mathbf v}) \rightarrow \mathbf t^* = - \frac{\nabla_{\mathbf v} v}{\vert\nabla_{\mathbf v} v\vert}
$$
this fact, in a different context, was recently successfully used by Izzo and Ozturk who used the HJB equations to directly inform the network training~\cite{dario_ekin_earth_venus}. The technique, developed independently but related to the emerging field of physically informed neural networks \cite{raissi2017physics}, made use of a training set of optimal trajectories defined in an ample bundle surrounding one nominal solution.
In general, when the aim is to  represent the optimal policy (i.e. the optimal low-thrust profile) in proximity to one nominal trajectory, for example for guidance and control purposes, the approach pursued in the previous section (i.e. a G\&CNET) is already able to produce very accurate results and while alternatives exist, they will not here be pursued.
Instead, we here seek to approximate the value function for the purpose of predicting the optimal rendezvous time-of-flight from any initial starting condition in a wide area of the entire state space, as opposed to targeting only a bundle around one nominal trajectory. In particular we focus here in a scenario where any orbit in an asteroid belt has to rendezvous to a target object placed in a circular, zero-inclined orbit.

\subsection*{The dataset}
We generate a dataset of optimal state-time-of-flight pairs by solving the shooting function in Eq.(\ref{eq:shooting}) for 4,000 different values of $\mathbf r_0, \mathbf v_0$. We generate these values by randomly sampling the osculating Keplerian elements in the inertial frame and then converting them into the Cartesian state in $\mathcal F$. The sampling bounds are chosen as to cover a good portion of the inner and middle asteroid belt with $a \in [2,3]$ AU, $e\in[0, 0.25]$, $i\in [0,5]$ deg. and $\Omega, \omega, E \in [0, 360]$ deg. In Figure \ref{fig:belt} the initial conditions are visualised as well as the target position for the rendezvous. 
The generic single entry in our dataset thus contains the values $[X_i, Y_i] = [(x_i,y_i,z_i,{v_x}_i,{v_y}_i,{v_z}_i), (t_{f_i}^*)]$. 
In this setup, the problem of local minima must be addressed with particular care as to avoid inserting in the dataset wrongful and deceiving information as well as avoiding database biases.
To this aim, we try to solve each of the 4,000 shooting functions in $N_{trials} = 120$ separate experiments , corresponding to the number of CPUs available to us. In each case, a maximum of $N_{res}=30$ restarts are tried to find one solution starting from randomly sampled co-states in $[-1,1]$ and  times-of-flight in $[1.5, 3.]$ years.
Out of the 120 guesses, if at least $N_{sol}=50$ result in a solution we then find, among them, the one corresponding to the smallest time of flight and insert it in the dataset, else we generate a new initial condition for the state and try again. We report that roughly 8\% of the initial conditions failed to meet this criteria (i.e. for those cases $N_{sol}<50$) and were thus substituted. Note that this kind of filter is likely to create a bias on the dataset towards simpler problems, as it discards the instances leading to a poor convergence of the root finder which may be also correlated to difficult cases for the regression, but the small fraction of discarded conditions was deemed to be acceptable for our purposes. We also report that in the creation of the dataset roughly 20\% of cases resulted in multiple locally optimal solutions (mostly two, of which we inserted in the dataset the one with smallest $t_f^*$) while the rest converged consistently (all $N_{trials}$ times) to the very same transfer. 
This procedure, although computationally expensive (taking less than half a day of computations on our supercomputer), allows to compile a dataset that includes correctly difficult cases for a low-thrust transfer such as those, for example, close to the hyper-surface of the state space separating trajectories requiring a different number of revolutions. 
These are in fact the conditions that often result in local minima or mis-convergence and that, if excluded, would result in a biased (and easier) dataset. If these conditions were instead included without caring for the possibility of local minima, the dataset would contain deceiving information for the model training. 
Eventually, we use 75\% trajectories for training and the remaining as a test set. 

We then produce also a second dataset, augmenting all the 3,000 training trajectories using the backward generation of optimal examples. 
For each nominal trajectory we produce 32 new optimal examples perturbing the final co-states by $\delta=0.0001$ and integrating back in time up to the optimal $t_f^*$. The resulting augmented dataset is created in less than 10 seconds on a standard CPU and contains 96,000 optimal transfers from the belt to the target. 

We release openly both dataset produced as well as the exact train/test split used in this manuscript on the CERN owned platform Zenodo~\footnote{The dataset can be downloaded here: \url{ https://doi.org/10.5281/zenodo.6365326} and was assigned the following d.o.i. \url{10.5281/zenodo.6365326} }.

\subsection*{The neural model}
We consider the task of learning the value function (i.e. the optimal time of flight) for the 1,000 trajectories of the test set as a supervised learning task, essentially a nonlinear regression problem. We use a simple feed-forward neural network with rectified linear units (ReLu) and 4 hidden layers with 100 neurons each. From the Cartesian elements $X$ contained in the dataset, we compute the modified equinoctial elements \cite{walker1985set}. In order to avoid issues related to the discontinuity of its representation, instead of the true longitude $L$, we use $\cos L, \sin L$ and we indicate the new set of attributes, now with an increased dimension, with $X_{eq}$. As a last step we scale all attributes uniformly in the range $[-1,1]$ via the simple transformation:
\begin{equation}
\begin{array}{l}
X_{in} = 2 (X_{eq} - X_{min})/(X_{max} - X_{min}) - 1\\
\end{array}
\end{equation}
For the optimizer we use the same setup (Adam \cite{kingma2014adam}) as in the experiments on the optimal policy, changing only the initial learning rate to $l = 0.5\cdot 10^{-3}$. 
Note that this regression problem, having a relatively small size, can also be tackled using different machine learning approaches (see for example \cite{mereta2017machine}) potentially offering improvements over the neural model here described. The neural approach here used, on the other hand, is of particular interest when coupled with the data augmentation technique employed, the backward generation of optimal examples, which creates a much larger training set to exploit.

\subsection*{Results and discussion}
\begin{figure}[tb]
\centering
 \includegraphics[width=1.0\textwidth]{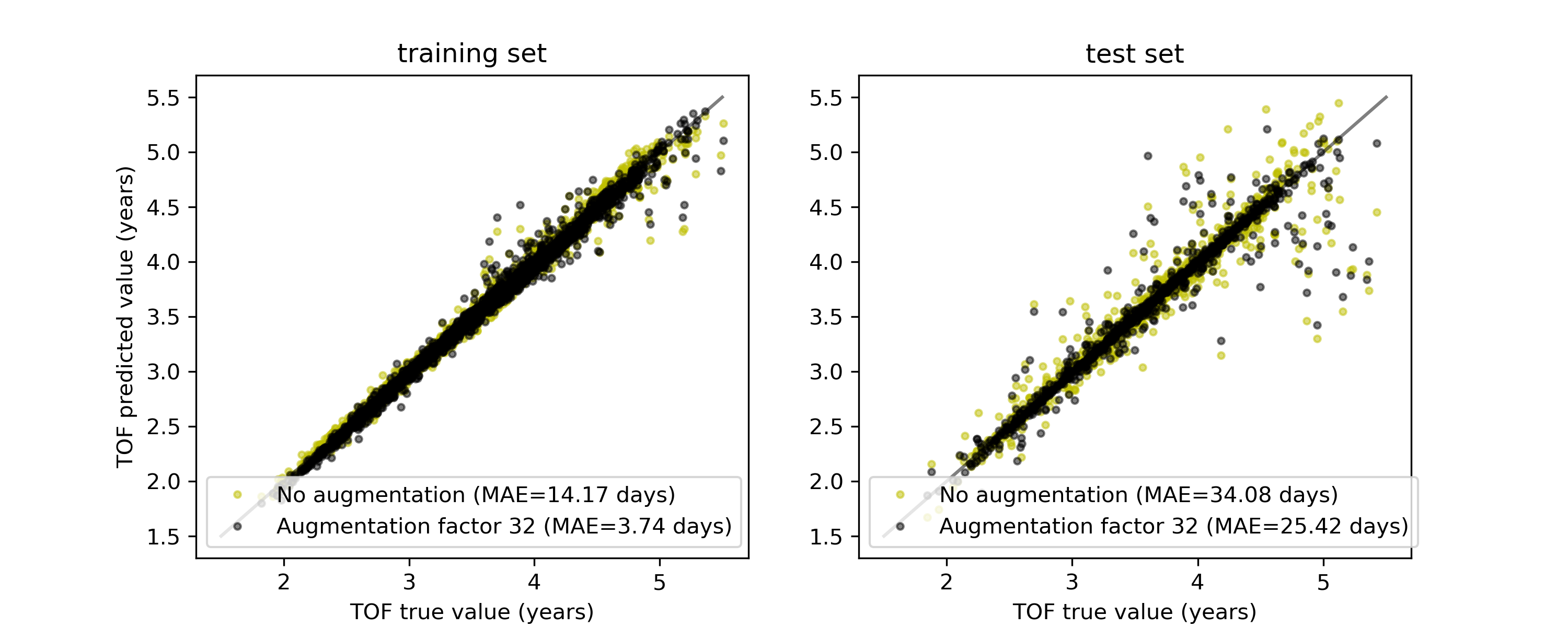}
\caption{Error on the training/test set.}
\label{fig:err_value_func_learn}
\end{figure}
\begin{figure}[tb]
\centering
\includegraphics[width=0.6\textwidth]{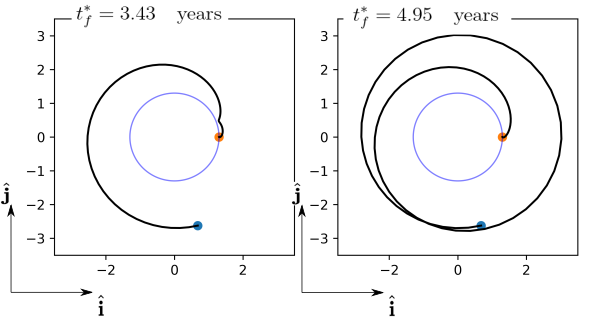}
\caption{An outlier in the neural model predictions over the test set. The globally optimal transfer is shown on the left, while the one consistently found by our numerical approach (and thus ending up in the dataset) is shown on the right.}
\label{fig:localglobal}
\end{figure}

The results for the best models trained over both the original and the augmented dataset are visualised in Figure \ref{fig:err_value_func_learn}. In general we note how, in relative terms, the error on the value function is now much larger than the one we obtained in a previous section when we trained the network to represent the optimal policy (even though still smaller than 4\% on average). 
This is expected and consequential to two different reasons. 
Firstly, we are tackling a much more difficult problem in that the diversity of the trajectories used is vastly larger, covering the entire asteroid belt, not only the neighbourhood of one nominal trajectory. 
Secondly, we are here learning from a much smaller and less dense dataset, a consequence of the fact that many nominal trajectories have here to be computed, not only one.
Nevertheless, the accuracy of $25.42$ days is greatly higher than, for example, the $468.56$ days one obtains using the Edelbaum approximation \cite{edelbaum1965optimum}.
This shows how the neural model developed is capable of capturing the complex phenomenon of phasing as well as effects connected to the orbits eccentricity and spatial orientation, aspects entirely neglected by the Edelbaum approach. 
A second fact worth noting is the advantage offered by the use of data augmentation. While all the augmented trajectories are in some neighbourhood of one entry of the original dataset, and thus add only a limited amount of new information, their presence allows for a significant improvement on the prediction (26\% decrease in MAE). It is worth mentioning here that a larger dataset, such as that produced by the data augmentation, also means higher training times for the model, something that would likely be accepted in a real application given the substantial gain introduced and the fact that the database augmentation cost, in terms of computational time is essentially, negligible.
On the right plot of Figure \ref{fig:err_value_func_learn} we see the generalization capabilities of the model on the test set. These are trajectories whose initial elements are still sampled uniformly in the same bounds as the training set, but that have never been encountered while training the model. 
It is interesting to note how very large errors are made by the model on a few trajectories, mostly in the test set, but also in the training set.
We thus performed an investigation on these outliers as to understand their origin. We find that in the test set there are 13 data points whose error on the predicted time of flight is larger than 300 days. For these 13 cases we find that the prediction of the trained neural model can be explained by one of three reasons. 
\begin{itemize}
    \item The predicted time of flight is actually correct and the label in the dataset, instead, correspond to a local minimum which slipped through our dataset creation pipeline. This actually happened for 6 of the 13 outliers: the majority of cases. This is particularly interesting as it hints at a possible use of the neural model to find or suggest possible cases where some full optimization model converges to a local minima.
    \item The predicted time of flight corresponds, within a few days, to a local minima. This actually happened for 2 of the 13 outliers. 
    \item The predicted time of flight does not represent any real optimal trajectory. This actually happened for 5 of the 13 outliers. While these cases are always bound to exist in the learned mode, removing erroneous labels in the training set and increasing the dataset size is likely beneficial to alleviate this issue.
\end{itemize}
In Figure \ref{fig:localglobal} we visualise one of the troublesome transfers, erroneously inserted in the dataset, but correctly predicted by the trained model, together with the correct globally optimal strategy. For this case, our root finder (applied to the shooting function) converged to the global solution only after many restarts, confirming this transfer as a particularly difficult case to find for any dataset creation procedure. One could of course try to mitigate this problem when creating a dataset by employing more restarts, but no number guarantees the absence of these occurrences. On the other hand, as here shown, the trained model does provide hints as to where they may occur, something that can be exploited \emph{a posteriori}.

\section*{Conclusion}
The time optimal low-thrust problem for a constant acceleration rendezvous has been studied in the context of neural modelling. 
Both the optimal policy and value function representation are studied.
We find that a G\&CNET, a deep feed forward neural network trained on optimal examples, is able to represent to great accuracy the optimal policy. 
Our results suggest that the on-board use of similar neural models for the spacecraft guidance and control is able to fulfil well possible operational requirements on the velocity and position residuals, thus constituting a valid alternative to currently proposed methods. 
We also show how the dataset of optimal trajectories employed to backpropagate the model errors can be created in seconds rather than days, if the data augmentation technique called backward generation of optimal examples is used.
In the case of value function learning we find that a neural model is able to represent the value function in a vast region of the state space corresponding to the entire asteroid belt and with errors that are below 4\% on average and mostly concentrated in a few outliers related to the presence of local minima solutions and, ultimately, of very complex globally optimal transfer strategies that compete with much simpler locally optimal ones.
We show that employing the data augmentation technique also in this case allows to significantly improve the predictions made by the neural model, leading to a 26\% decrease in MAE.
While the dynamical model used in this paper as a test case is not appropriate for many cases of practical interest, it served its purpose of resulting in quick optimisation procedures and thus allowing to refine the overall work logic and results. We hope that our work will facilitate, in the future, the training of neural models representing more complex cases.

\bibliography{sample}

\end{document}